\documentclass[conference,13]{IEEEtran}
\IEEEoverridecommandlockouts
\usepackage{cite}
\usepackage{amsmath,amssymb,amsfonts}
\usepackage{algorithmic}
\usepackage{graphicx}
\usepackage{textcomp}
\usepackage{xcolor}
\usepackage{tikz}
\usepackage[nodisplayskipstretch]{setspace}
\usepackage{glossaries} 

\def\BibTeX{{\rm B\kern-.05em{\sc i\kern-.025em b}\kern-.08em
    T\kern-.1667em\lower.7ex\hbox{E}\kern-.127emX}}

\makeatletter
\newcommand{\linebreakand}{%
  \end{@IEEEauthorhalign}
  \hfill\mbox{}\par
  \mbox{}\hfill\begin{@IEEEauthorhalign}
}
\makeatother

\newacronym{DES-POMDP}{DEC-POMDP}{decentralized partially observable Markov decision process}
\newacronym{DLT}{DLT}{distributed ledger technology}
\newacronym{MARL}{MARL}{multi-agent reinforcement learning}
\newacronym{MNO}{MNO}{mobile national operator}
\newacronym{PPO}{PPO}{proximal policy optimization}
\newacronym{QoS}{QoS}{quality of service}
\newacronym{RAN}{RAN}{radio access network}
\newacronym{RL}{RL}{reinforcement learning}
\newacronym{SD}{SD}{social dilemma}
\newacronym{SP}{SP}{service provider}
\newacronym{UE}{UE}{user equipment}

\begin{document}

\title{Evaluating Inter-Operator Cooperation Scenarios to Save Radio Access Network Energy}

\author{

\IEEEauthorblockN{Xavier Marjou}
\IEEEauthorblockA{\textit{Orange Labs Lannion} \\
\textit{Orange}\\
Lannion, France \\
xavier.marjou@orange.com}
\and
\IEEEauthorblockN{Tangui Le Gleau}
\IEEEauthorblockA{\textit{Orange Labs Lannion} \\
\textit{Orange}\\
Lannion, France \\
tangui.legleau@orange.com}
\and
\IEEEauthorblockN{Vincent Messie}
\IEEEauthorblockA{\textit{Orange Labs Lannion} \\
\textit{Orange}\\
Lannion, France \\
vincent.messie@orange.com}
\linebreakand 
\IEEEauthorblockN{Benoit Radier}
\IEEEauthorblockA{\textit{Orange Labs Lannion} \\
\textit{Orange}\\
Lannion, France \\
benoit.radier@orange.com}
\and
\IEEEauthorblockN{Tayeb Lemlouma}
\IEEEauthorblockA{\textit{IUT Lannion} \\
\textit{IRISA}\\
Lannion, France \\
tayeb.lemlouma@irisa.fr}
\and
\IEEEauthorblockN{Gael Fromentoux}
\IEEEauthorblockA{\textit{Orange Labs Lannion} \\
\textit{Orange}\\
Lannion, France \\
gael.fromentoux@orange.com}
}

\maketitle

\begin{abstract}
Reducing energy consumption is crucial to reduce the human debt's with regard to our planet. Therefore most companies try to reduce their energetic consumption while taking care to preserve the service delivered to their customers. To do so, a \gls{SP} typically downscale or shutdown part of its infrastructure in periods of low-activity where only few customers need the service. However an SP still needs to maintain part of its infrastructure "on", which still requires significant energy. For example a \gls{MNO} needs to maintain most of its \gls{RAN} active. Could an \gls{SP} do better by cooperating with other \gls{SP}s who would temporarily support its users, thus allowing it to temporarily shut down its infrastructure, and then reciprocate during another low-activity period? To answer this question, we investigated a novel collaboration framework based on \gls{MARL} allowing negotiations between \gls{SP}s as well as trustful reports from a \gls{DLT} to evaluate the amount of energy being saved. We leveraged it to experiment three different sets of rules (free, recommended, or imposed) regulating the negotiation between multiple \gls{SP}s (3, 4, 8, or 10). With respect to four cooperation metrics (efficiency, safety, incentive-compatibility, and fairness), the simulations showed that the imposed set of rules proved to be the best mode.

\end{abstract}


\section{Introduction}

Reducing energy consumption is crucial to reduce human debt to our planet. Many governments and enterprises have taken action plans to significantly reduce their energy consumption in the coming years. Thus most services providers (\gls{SP}s) strive to reduce their consumption by taking care not to impact the service delivered to their customers (e.g. \cite{b1}, \cite{b8}). Typically, in periods of low-activity when only few customers need the service, they scale down or shutdown part of their infrastructure. For instance, during some low traffic hours at night, mobile network operators (\gls{MNO}s) generally have few active subscribers and thus may shut-down some frequency bands while only keeping alive one frequency band, usually the lowest-frequency band to provide cellular connectivity in a maximal geographical area (e.g. \cite{b11}).

However, most SPs still need to maintain part of their infrastructure on, which still requires significant energy. For instance, \gls{MNO}s need to keep most part of their radio access network (\gls{RAN}) active despite having few customers because the \gls{RAN} needs to remain immediately reachable by any subscriber's \gls{UE}.

In order to bring energy savings to a higher level, an uttermost solution might be to collaborate among SPs delivering the same type of service. Provided that an SP could serve all the users of others SP over a low-activity period, it could offer an \emph{on-guard} service to serve the users of partner \gls{SP}s during a low-activity period. For instance, an \gls{MNO} could allow the other \gls{MNO}s' users to access its network via national roaming during this period. In exchange, the partner \gls{SP}s might be on-guard during subsequent low-activity periods, allowing the \gls{SP} to temporarily shutdown its whole infrastructure and achieve significant energy savings. However, would \gls{SP}s cooperate? And if so, under which environmental constraints? Should there be no constraint so that \gls{SP}s would freely interact, the cooperation dynamic nevertheless emerging naturally? Or should a supervisor at least recommend the \gls{SP} to be on-guard? Or should a supervisor go even further and impose the \gls{SP} to be on-guard?

To answer this question, we experimented a novel tit-for-tat cooperative framework integrating a central on-guard service, with a \emph{negotiation} interface for on-guard service offer and demand between \gls{SP}s, as well as a \emph{reporting} interface to a distributed ledger technology (\gls{DLT}), as shown in Fig. \ref{fig:architecture}. We implemented the service as a \gls{RL} environment so that simulated agents, each representing an \gls{SP}, could use an off-the-shelf \gls{RL}-based policy to rationally interact with each other through the service which acts as an intermediary. The service could be configured and launched in one of the following modes: \emph{free}, \emph{recommended} or \emph{imposed}; and it could also access trustful reports from the \gls{DLT} to totalize the kWh being consumed by the on-guard \gls{SP}.

We tested 12 independent scenarios: each of the three modes was tested with a set of 3, 4, 8, or 10 \gls{SP}s. For each scenario, each \gls{SP} used \gls{PPO} \cite{b4} as its \gls{RL} policy; after training the \gls{SP}s' policies, we evaluated four cooperation metrics (efficiency, safety, incentive-compatibility and fairness) to evaluate the outcome of the scenario from an oracle perspective.

Among the tested modes, the results showed that the \emph{imposed} scenario proved to be the best to meet the best cooperation properties among participating \gls{SP}s.

A limitation of our work is that the \gls{QoS} of the service provided during the collaboration was not taken into account, which could lead to tensions among the collaborative \gls{SP}s in case the QoS is not acceptable to customers. We thus envisage to extend the framework with respect to QoS to enhance \emph{trust}, especially in a use-case involving \gls{MNO}s.

Another limitation is related to the key criteria of the agent's policy. To ensure that the policy is efficient, safe, incentive-compatible and fair, additional work would be needed to ensure that a policy like \gls{PPO} is robust against a variety of other policies that other \gls{SP}s might use.

We believe that our findings are important to save energy because the cooperation is not limited to two \gls{SP}s. On the contrary, the more cooperating \gls{SP}s there are, the greater the energy savings are for each \gls{SP} given that the order of magnitude of the energy savings is $(N-1)/N$ for each of the $N$ cooperating \gls{SP}s, which should be compelling enough to gain attention.

\section{Related Works}

Axelrod's tit-for-tat \cite{b10} is a key algorithm regarding cooperation, as it is optimal on several aspects: in addition to be \emph{efficient} with regard to the social welfare (sum of the utilities for each participant), it is also \emph{secure} in the sense that a participant does not need to be afraid of being exploited, and \emph{incentive-compatible} in the sense that its participation will encourage the other participant to play. In addition, the algorithm is \emph{provocable}: it stops cooperating upon detecting that the other participant has stopped cooperating; and is \emph{forgivable}: it starts again cooperating upon detecting that the other participant cooperates again. However, the tit-for-tat algorithm is implemented in the agent's side (i.e., the task to be done by the \gls{SP}), which does not allow to cope with scenarios where an intermediary would be involved to possibly influence, slightly or strongly, the negotiation between the agents.

The work of Lerer and Peysakhovich \cite{b9} brought considerable innovation introducing \gls{RL} to lead cooperation between two agents. They avoid being trapped in a social dilemma by iterating interactions and finally act in ways that are simple to understand, nice (begin by cooperating), provocable (try to avoid being exploited), and forgiving (try to return to mutual cooperation). 

More recently, Zheng et all \cite{b7} studied whether multiple agent reinforcement learning (\gls{MARL}) could bring equality and productivity with AI-driven tax policies with a inner-outer loop. The inner loop allowed the agents to learn to maximize their utility by performing labor, receiving income, and paying taxes, while the outer loop allowed the intermediary (a.k.a. the regulator) to optimize taxes for any social objective. Though being extremely promising, this work does not seem to allow a read-only participation from the intermediary, which is one important mode to be evaluated.

\section{Model}

\subsection{Social Dilemma}
\label{sec:social_dilemma}

\begin{figure}
    \centering
    
\begin{tikzpicture}

\tikzstyle{label} = [minimum width=0.2cm, minimum height=0.7cm,text centered]
\tikzstyle{symbol} = [minimum width=0.35cm, minimum height=0.7cm,text centered]
\tikzstyle{arrow} = [-stealth,->]

\node (sp0) [symbol] {Service Provider 1};
\node (sp1) [symbol,below of=sp0, xshift=0cm, yshift=-0.1cm] {Service Provider i}; 
\node (sp2) [symbol,below of=sp1, xshift=0cm, yshift=-0.1cm] {Service Provider N}; 
\node (esp) [symbol,right of=sp1, xshift=2.5cm,align=center]{On-guard Service};
\node (reg) [symbol,above of=esp, xshift=0cm, yshift=+0.4cm] {Regulator};
\node (dlt) [symbol,below of=esp, xshift=0cm, yshift=-0.4cm] {Distributed Ledger};
\draw [arrow] (sp0) -- (esp);
\draw [arrow] (sp1) -- (esp);
\draw [arrow] (sp2) -- (esp);
\draw [arrow] (esp) -- (sp0);
\draw [arrow] (esp) -- (sp1);
\draw [arrow] (esp) -- (sp2);
\draw [arrow] (reg) -- (esp);
\draw [arrow] (dlt) -- (esp);

\end{tikzpicture}
\caption{Architecture}
\label{fig:architecture}
\end{figure}
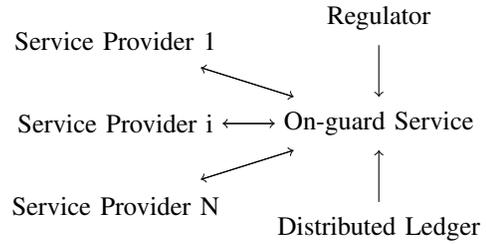

To give an intuition on how \gls{SP}s may consider cooperating to save energy on low-activity periods, let's model it as a \gls{SD} with the following costs associated to the three possible states of the \gls{SP}'s infrastructure: 
\begin{itemize}
  \item $o$ when the \gls{SP}'s infra is off;
  \item $a$ when the \gls{SP}'s infra in active for its users only;
  \item $g$ when the \gls{SP}'s infra is on-guard for any user.
\end{itemize}
The matrix of Table \ref{tab:isd}  represents the possible \emph{cooperate} (C) and \emph{defect} (D) actions and payoffs when the \gls{SD} is iterated twice (i.e., two consecutive negotiation cycles). Note that this representation is limited to two agents and that we only detail the four most relevant payoff values.

\begin{table}[h!]
\caption{A two-player iterated social dilemma after two negotiations.}
\begin{center}
\begin{tabular}{ |c|ccc|ccc| } 
 \hline
   & &C1& & &D1& \\ 
 \hline
  & & C2 & D2 & & C2& D2\\
 C1 & C2 & $R=g+o$ & ... & C2 & ... & $S=g+g$ \\ 
  & D2 & ... & ... & D2 & ... & ... \\ 
 \hline
  & & C2 & D2 & & C2& D2\\
 D1 & C2 & ... & ... & C2 & ... & ... \\ 
  & D2 & $T=o+o$ & ... & D2 & ... & $P=a+a$ \\ 
 \hline
\end{tabular}
\end{center}
\label{tab:isd} 
\end{table}

By setting
\begin{itemize}
  \item $o=-0.01$ (the \gls{SP}'s infra  set to off costs nearly nothing)
  \item $a=-0.90$  (the \gls{SP}'s infra  maintained on is expensive)
  \item $g=-1.00$ (the \gls{SP}'s infra  maintained on for all users)
  \end{itemize}
the payoffs become T= -0.02, R = -1.01, P=-2.00, S=-1.80.

Hence, with such settings, this resulting dilemma leads to a prisoner's dilemma as $T>R>P>S$ (cf. \cite{b10}). In this specific case, it is unknown if cooperation can emerge across iterations since $2*R = T+S$.

\subsection{Architecture}

To cope with scenarios involving the negotiation between multiple \gls{SP}s as well as the possible intervention of the regulator, we used the architecture depicted in Fig. \ref{fig:architecture}. All negotiation messages between the participating \gls{SP}s go through a centralized on-guard service. 

The service can be configured and observed by the regulator (e.g. to set specific rules, to provide an estimation of the energy consumed by each \gls{SP}, or to configure a blacklisting policy, if any, as well as its duration).

In addition, an interface to a Distributed Ledger (\gls{DLT}) reports \emph{a posteriori} the energy (kWh) officially consumed by the \gls{SP} being on-guard during the previous low-activity period. The service may thus update the total amount of kWh consumed by each \gls{SP} in order to calculate the fairest sharing between \gls{SP}s and possibly recommend or impose the \gls{SP} to be on-guard for the next low-activity period. In the future, the DLT could also integrate other metrics for measuring the \gls{QoS} of the service provided by the on-guard \gls{SP}. 

\subsection{Environmental modes}

The service can be instantiated in one of the following \emph{modes}: 
\begin{itemize}
  \item \emph{Free} (F): the service makes no suggestion as to which \gls{SP}(s) should be on-guard and act as a pure relay with regard to the \gls{SP}s' actions; it lets the \gls{SP}s decide who is/are the on-guard \gls{SP}(s) for the next low-activity period.
  \item \emph{Recommended} (R): at the initialization time of each new negotiation, the service suggests in the observation message which \gls{SP} should be on-guard, selecting the \gls{SP} that brought the lowest total amount of saved energy for the other \gls{SP}s.
  \item \emph{Imposed} (I): in addition to suggesting which \gls{SP} should be on-guard, the service also enforces offers coming exclusively from that \gls{SP}. But it does not impose an \gls{SP} to offer on-guard service to all demanding \gls{SP}s. The service also temporarily blacklists an \gls{SP} that has refused to be on-guard more than $m$ times and can reintegrate it after a blacklisting period is finished. As a consequence, this mode takes advantage of the service's central intermediary position to enforce \emph{provocability} and \emph{forgiveness} on behalf on the participating \gls{SP}s; it also simplifies the strategy and makes it \emph{clear} by announcing and imposing the \gls{SP} to be on-guard. Note that these three properties are instead performed decentrally in \cite{b10}.
\end{itemize}

\subsection{Negotiation interface}

The negotiation interface allows for actions from the \gls{SP}s to the service. Each negotiation involves a set of messages between \gls{SP}s to determine the \gls{SP}, or the \gls{SP}s, that will share its/their infrastructure for the next low-activity period.

The messages from an \gls{SP} to the service contain a single action parameter containing a set of sub-actions. There are two types of sub-actions: an \emph{offer} so that a $\gls{SP}_i$ can offer another $\gls{SP}_{j, j \neq i}$ to be on-guard (i.e,. offer to serve $\gls{SP}_j$'s customers) and a \emph{demand} so that a $\gls{SP}_i$ can demand another $\gls{SP}_{j, j \neq i}$ to be on-guard (i.e., demand to serve $\gls{SP}_i$'s customers)

The messages from the service to each \gls{SP} contain an \emph{observation} of the offers and demands from the other \gls{SP}s to this \gls{SP}, as well as a positive or negative \emph{reward} proportional to the energy estimated to be saved or consumed during the next low-activity period based on the results of the current negotiation plus a small negative reward at each timestep in order to speed the negotiation.

\subsection{POMDP}

We consider N players, where each player represents an \gls{SP}. The players repeatedly interact with each others accross multiple negotiations, with each negotiation lasting multiple timesteps. 

We formulate the negotiations as a variant of a \gls{DES-POMDP} \cite{b12} with individual rewards, thereby allowing choices to be made decentrally by the set of agents. Our variant is defined by an tuple $\left\langle\mathcal{I}, \mathcal{S}, \mathcal{A},O,\mathcal{T}, R, \gamma \right\rangle$ where $\mathcal{I} = \{1,..., N\}$ is a set of agents, $\mathcal{S}$ is the set of states and $O : \mathcal{S} \times \mathcal{I} \rightarrow \mathcal{S}$ is an observation function. $\mathcal{A} = \mathcal{A}_1 \times ... \times \mathcal{A}_N$ is the set of joint actions, and a joint action $\vec{a} = (a_1, ..., a_N)$ transitions the state $s$ to the next state following the stochastic function $\mathcal{T} : \mathcal{S} \times \mathcal{A}_1 \times ... \times \mathcal{A}_N \rightarrow \Delta(\mathcal{S})$. At last, a personal reward function $R : \mathcal{I} \times \mathcal{S} \times \mathcal{A} \rightarrow \mathbb{R} $ gives a reward $r^i$ to each player $i$. The joint reward is denoted $\vec{r} = (r^1,...,r^N)$. Each agent$_i$'s goal is to find a policy $\hat\pi_i : \mathcal{S} \rightarrow \Delta(\mathcal{A}^i)$ in order to maximise its expected discounted return defined by:
\[ G_{\hat{\pi}_i}(s_0) = \mathbb{E}[\sum_{t=0}^{\infty} \gamma^t r^i(s_t,\vec{a}_t) | \vec{a}_t \sim \hat{\pi}_i, s_{t+1} \sim \mathcal{T}(s_t, \vec{a}_t)   ]   \]
$\vec{\pi}$ denotes the joint policy $(\hat\pi_1,...,\hat\pi_N)$.

\section{Experimentation}

\subsection{Environment}

The service was implemented as an RL environment that can be instantiated in one of the three modes (F, R or I). Each environment's episode corresponds to a negotiation where the participating agents exchange messages to determine the on-guard agent for the next low-activity period. 

For all modes, we used the same reward values: $-1.00$ when an agent successfully negotiated to be on-guard; $-0.9$ when an agent was not able to successfully find an on-guard \gls{SP}; $-0.01$ when an agent successfully found an on-guard \gls{SP}; and $-0.01$ for each other time-step to incite the agents to use the smallest number of messages during an episode. We set the maximal number of steps per episode to $10$ in order to enforce short negotiations. 

We implemented the environment as a Gym environment \cite{b2}, a Python toolkit for developing and comparing RL algorithms. We used a multi-discrete binary value for the action space and also a multi-discrete binary value for the observation space, in order to faster the convergence of the RL training.

Regarding the interface between the environment and the \gls{DLT}, we simulated energy consumption reports with a function adding a 10\% random noise to the kWh estimated to be consumed by the on-guard \gls{SP}.  

\subsection{Agents}

We implemented the agents with RLlib \cite{b3}, a Python library allowing for scalable MARL. 

Regarding the RL policy algorithm, we selected Proximal Policy Optimization (\gls{PPO}) for all agents as some of our pre-experiments showed that \gls{PPO} \cite{b4} performed better than other algorithms (PG, A2C, A3C or MARWIL), which has also been successfully used in other \gls{MARL} experiments \cite{b13}. Future work may experiment agents with different policies.

\subsection{Cooperation properties}
\label{sec:metrics}

\newcommand{\cstNorm}[0]{I_{norm}}

To evaluate the cooperation across episodes, we introduced four properties \cite {b14}, also known as societal objectives \cite{b15}. Our three first metrics were adapted from \cite{b9} whereas the fourth one is the Jain's fairness index of \cite{b6}.

After a training step $r$ involving $N$ agents, each agent $i$ acting on the behalf of one \gls{SP} and performing actions based on its learned \gls{RL} policy $\hat{\pi_{i}}$, negotiate the on-guard service during $T_{max}$ episodes. $G_i(\hat{\pi}, t)$ is the expected return, i.e. the sum of of discounted rewards received by an agent $i$ during one episode $t$ (i.e., one negotiation). 

An \emph{efficiency} property ($E$) measures how close the social welfare is from the optimum, which happens when all agents cooperate. The social welfare is the sum of expected returns over all agents. An optimal cooperating policy is noted $\pi_{i,C}$, whereas the worst cooperating policy is noted $\pi_{i,D}$. 

\begin{center}
\[
    E(\vec{\pi}, r, t) = \frac{\sum_{i=1}^{N}G_i(\hat{\pi}_i, r, t) - \sum_{i=1}^{N}G_i(\pi_{i,D}, r, t)}{\sum_{i=1}^{N}G_i(\pi_{i,C}, r, t) - \sum_{i=1}^{N}G_i(\pi_{i,D}, r, t)}
\]
\end{center}

A \emph{safety} ($Sf$) property measures the risk taken by an agent $i$ and is defined as the difference, when all other agents defect ($\pi_{-i,D}=\pi_{j, j \neq i, D}$), between the expected return received by agent $i$ when trying to cooperate and the expected return it received when defecting. $(T - S)$ represents the maximal amplitude (cf. Section \ref{sec:social_dilemma}) and, together with $+1$, allow to normalize $Sf$ between $0.0$ and $1.0$.

\begin{center}
\[
    Sf(\hat{\pi}_i, r, t) = \frac{G_i(\hat{\pi}_i, \pi_{-i,D} , r, t) - G_i(\pi_{iD}, \pi_{-i,D}, r, t)}{T-S} + 1
\]
\end{center}

An \emph{incentive-compatibility} property ($IC$) measures the capacity to incentivize cooperation and is defined as the difference, when all other agents try to cooperate ($\hat{\pi}_{-i}=\hat{\pi}_{j, j \neq i}$), between the expected return received by an agent $i$ when trying to cooperate and the expected return it received when defecting. $(T - S)$ represents the maximal amplitude (cf. Section \ref{sec:social_dilemma}) and allows to normalize $IC$. 

\begin{center}
\[
    IC(\hat{\pi}_i, r, t) = \frac{G_i(\hat{\pi}_i, \hat{\pi}_{-i}, r, t) - G_i(\pi_{i,D}, \hat{\pi}_{-i}, r, t)}{T -S}
\]
\end{center}

Finally, a \emph{fairness} property (J) with $e_i$ equal to the total number of kWh saved by each agent at the end of an episode $t$ since the beginning of the evaluation after the training. In case of blacklisted agent(s), the Jain index was calculated based on the non-blacklisted agents.

\begin{center}
\[
    J(\vec{\pi}, r, t)=\frac{(\sum_{i=1}^{N}e_{i,t,r})^2}{N\sum_{i=1}^N {e_{i,t,r}}^2}
\]
\end{center}

\subsection{Training and metrics measurement}

We performed experiments with four different numbers of agents (N = 3, 4, 8, or 10) and three different modes on the environment side (M = F, R, or I).

For each (N, M) experiment, we first performed a set of 5 independent training runs. Each run was stopped when plateauing during 20000 timesteps. For each run, we launched 100 episodes and measured the cooperation properties by an oracle implemented in the environment. We then measured the distribution (mean and standard deviation) of these properties across the $5*100$ episodes. 

We limited the maximum number of agents to 10 because one important use-case assumption is that there should be only one on-guard \gls{SP} and that it must have the possibility to host all the users of all \gls{SP}s, which prevents experimenting a big number of agents. Another reason is that each training time increases quadratically with the number of participating agents. Future work may involve a game allowing for multiple on-guard \gls{SP}s and thus a greater number of agents.

\section{Results and Discussion}

\subsection{Analysis of the scenarios}

\begin{figure}[h!]
\begin{center}
\begin{tabular}{cc}
\includegraphics[width=3.75cm]{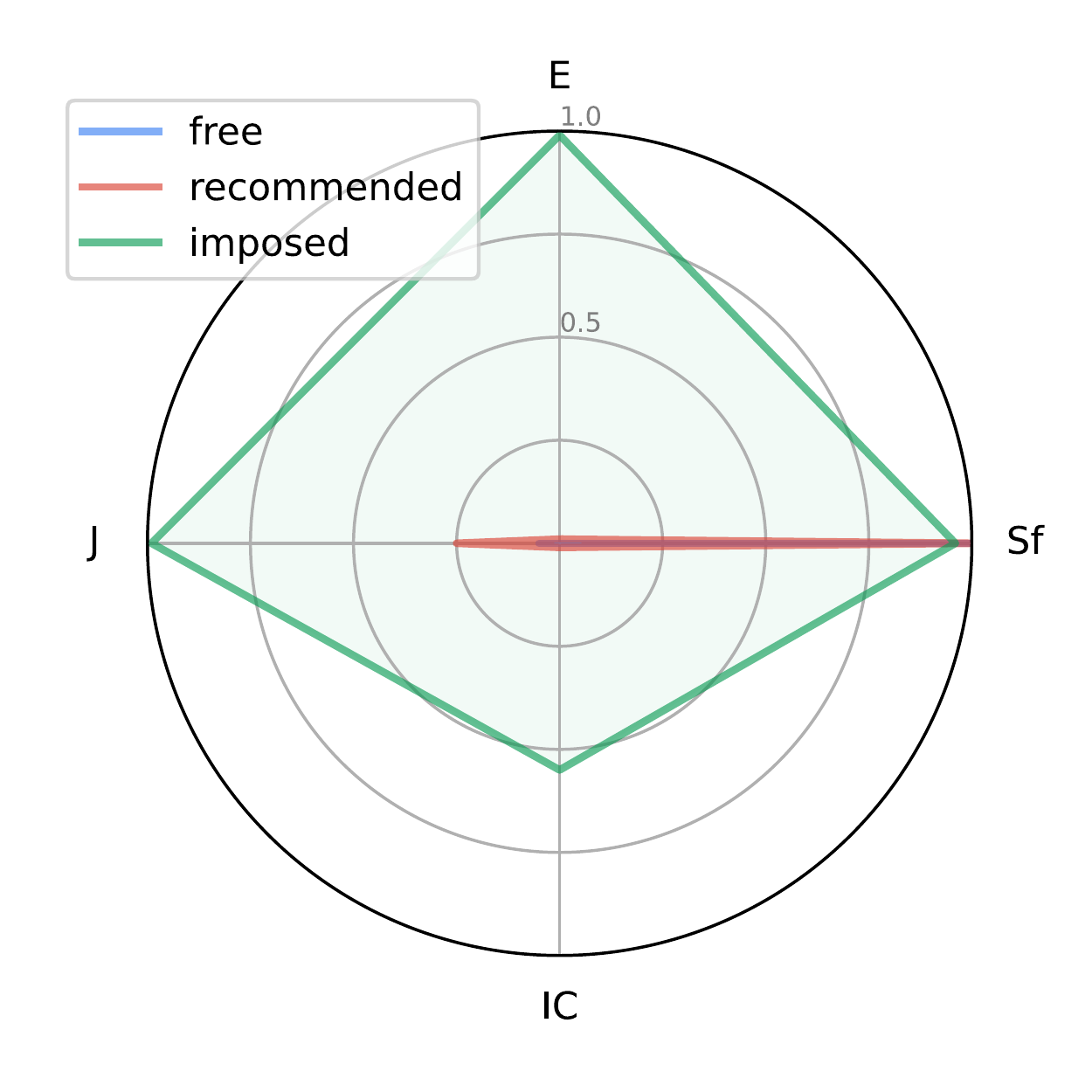} &
\includegraphics[width=3.75cm]{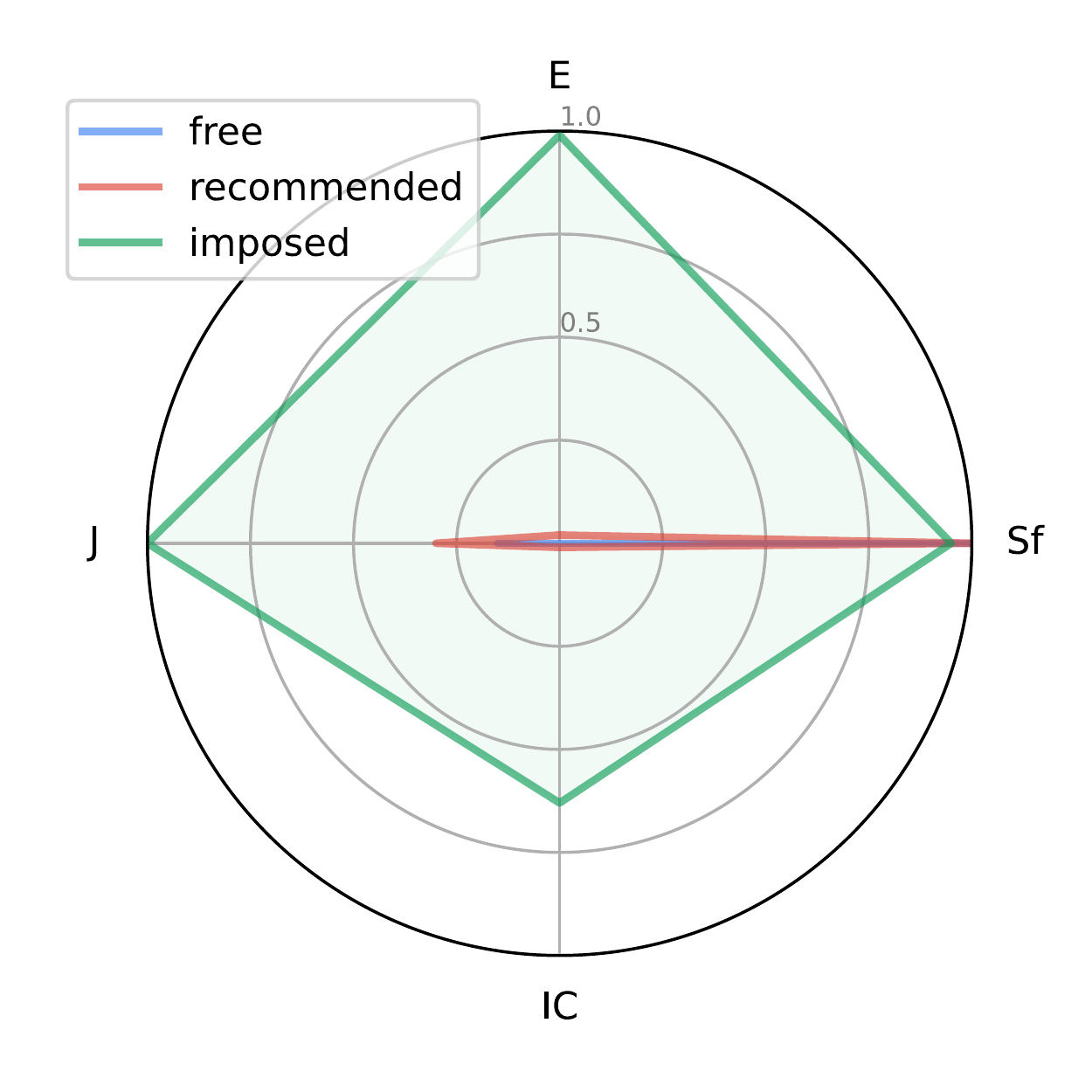}\\
\small{3-agent game} & \small{4-agent game} \\
\includegraphics[width=3.75cm]{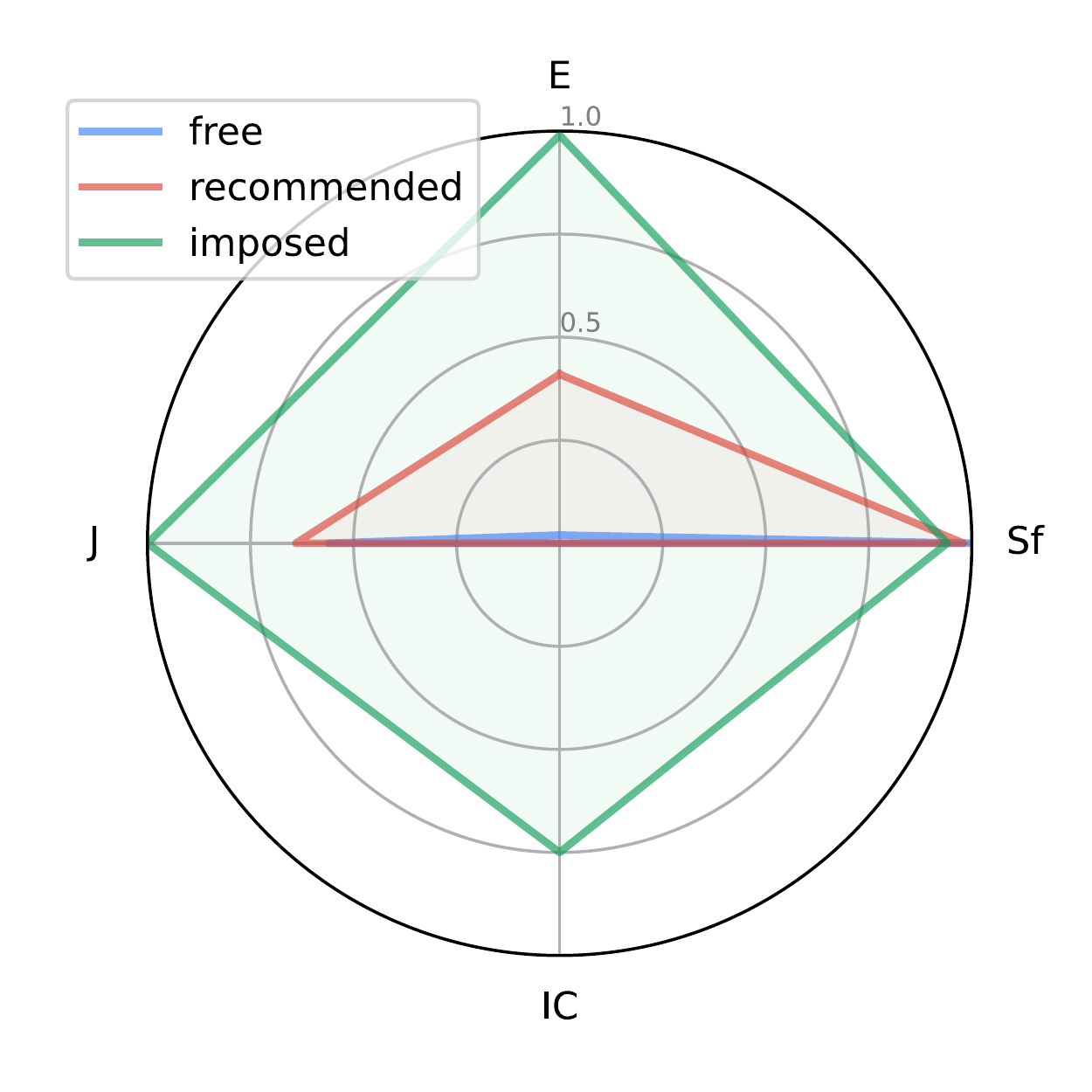} &
\includegraphics[width=3.75cm]{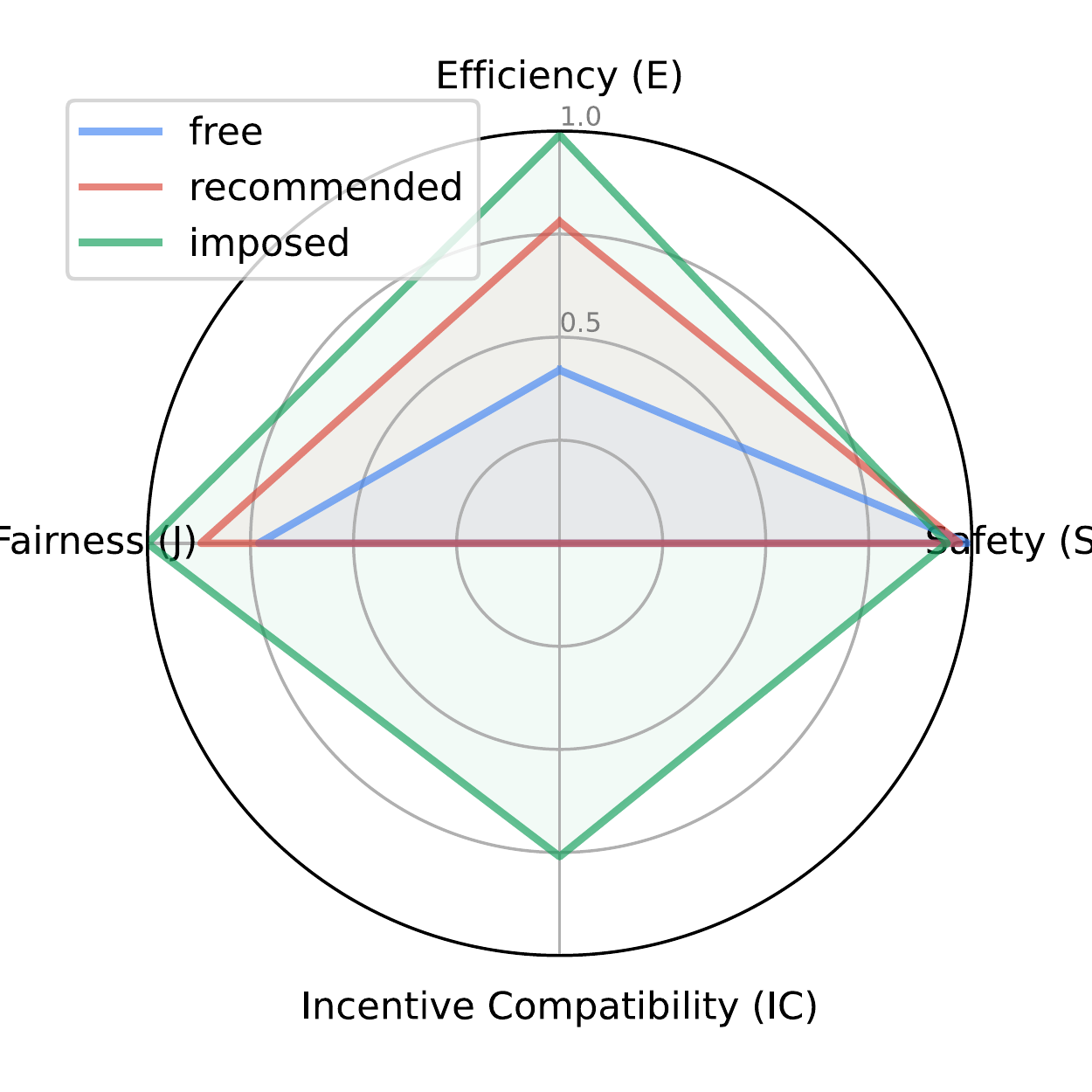}\\
\small{8-agent game} & \small{10-agent game} \\
\end{tabular}

    \caption{Radar chart of the cooperation metrics}
    \label{fig:radar_chart} 
\end{center}
\end{figure}

Table \ref{tab:cooperation_metrics_results} and Fig. \ref{fig:radar_chart} display the obtained results.

\begin{table}[h!]
\tiny{
\small
  \caption{ Efficiency (E), Safety (Sf), Incentive-compatibility (IC) and Fairness (J) properties according to the number of participating agents (N) and to the tested mode (M = free (F), or recommended (R) or imposed (I))}
    \label{tab:cooperation_metrics_results}
  \centering
  \tiny{}
\begin{center}
\begin{tabular}{ |c|c|c|c|c|c| }
\hline
N & M & E & Sf & IC & J \\
\hline
3 & F & $0.00\pm0.02$ & $1.00\pm0.00$ & $0.00\pm0.08$ & $0.05\pm0.12$ \\
3 & R & $0.00\pm0.04$ & $1.00\pm0.01$ & $0.00\pm0.06$ & $0.25\pm0.24$ \\
3 & I & $\boldsymbol{0.99\pm0.03}$ & $\boldsymbol{0.96\pm0.04}$ & $\boldsymbol{0.55\pm0.44}$ & $\boldsymbol{0.99\pm0.03}$ \\
\hline
4 & F & $0.00\pm0.04$ & $1.00\pm0.01$ & $0.00\pm0.09$ & $0.15\pm0.19$ \\
4 & R & $0.01\pm0.07$ & $1.00\pm0.01$ & $0.00\pm0.09$ & $0.30\pm0.28$ \\
4 & I & $\boldsymbol{0.99\pm0.02}$ & $\boldsymbol{0.95\pm0.04}$ & $\boldsymbol{0.63\pm0.40}$ & $\boldsymbol{1.00\pm0.03}$ \\
\hline
8 & F & $0.02\pm0.08$ & $1.00\pm0.00$ & $0.00\pm0.17$ & $0.56\pm0.33$ \\
8 & R & $0.41\pm0.47$ & $0.98\pm0.02$ & $0.00\pm0.23$ & $0.64\pm0.24$ \\
8 & I & $\boldsymbol{0.99\pm0.01}$ & $\boldsymbol{0.94\pm0.04}$ & $\boldsymbol{0.75\pm0.31}$ & $\boldsymbol{1.00\pm0.01}$ \\
\hline
10 & F & $0.42\pm0.47$ & $0.99\pm0.02$ & $0.00\pm0.24$ & $0.73\pm0.23$ \\
10 & R & $0.78\pm0.37$ & $0.97\pm0.03$ & $0.00\pm0.23$ & $0.87\pm0.08$ \\
10 & I & $\boldsymbol{0.99\pm0.01}$ & $\boldsymbol{0.94\pm0.04}$ & $\boldsymbol{0.76\pm0.28}$ & $\boldsymbol{1.00\pm0.01}$ \\
\hline
\end{tabular}
\end{center}}

\end{table}

In most scenarios with \emph{free} modes, the efficiency was close to $0.0$. The incentive to play was also close to $0.0$: if an \gls{SP} tried to cooperate with other cooperating \gls{SP}s, its expected return did not increase;  However the safety remained closed to $1.0$, suggesting that an \gls{SP} did not take many risks by trying to cooperate. The results led to unfair fairness scores, although being calculated on the rare successful transactions.

In all scenarios with \emph{imposed} modes, E remained close to $1.0$. Recall that when E is close to $1.0$, each of the N agents saves $N-1/N$ of energy per low-activity period. Thus, as energy savings increase with N, a maximum number of cooperating \gls{SP}s is nevertheless desirable. This can also be noticed with the values of IC: indeed, when all but one agent collaborate, the more agents there are, the bigger the regret for the defecting agent. In addition, the results also showed excellent results for the safety and fairness scores. Such successful and reiterated cooperations suggest that a strong control from a regulator can be beneficial for all agents, provided that this control is aligned with convincing rewards for the agents. Otherwise the agents would likely have no incentive to cooperate and would stop performing offers.

In scenarios with \emph{recommended} modes, there were two subsets of results. With N=3 or N=4 agents, the mean efficiency remained close to zero, as in the free modes; similarly, the incentive-compatibility was also close to zero and the safety remained high indicating no risk to play. Instead, with N=8 or N=10 agents, the recommended modes demonstrated more and more efficiency, although being subject to important variance across different runs and although fairness remained limited. This suggests that recommending the agent to be on-guard in the observation is a useful hint for most agents but is not sufficient to trigger a virtuous equilibria among multiple agents; The fact that efficiency sometimes emerged when $N\geq8$ might be due to the few agents who have early discovered cooperation by chance and who have progressively influenced the behavior of other agents. Consequently, future work should strive to investigate the reasons for the successful cases and identify whether cooperation could be stabilized with additional hints or different reward, which, if successful, would result in a smoothly controlled mode would be very appealing.

A limitation of this work is that the \gls{QoS} provided during the cooperation is not taken into account, which could lead to tensions among \gls{SP}s in case of low \gls{QoS}. As a consequence, future work should also integrate \gls{QoS}.

Note: These results are purely technical and do not presume whether any of the mode would be allowed by a real regulator.

\section{Conclusion}

We described a trusted and collaborative framework based on \gls{MARL} allowing a regulator to test different rules in order to supervise negotiations related to energy savings, such as \gls{RAN} energy savings, and also allowing service-providers to train and evaluate their policy regarding such a participation.

The analysis of cooperation metrics showed that successful cooperation emerged upon tightly constrained rules enforcing the scheduling of the \gls{SP} to be on-guard. However cooperation hardly emerged from loosely constrained rules, although there might be room for improvement. 

As such, we believe that this framework might be beneficial to evaluate cooperative scenarios to significantly reduce energy consumption.

\end{document}